\documentclass[12pt]{article}%
\usepackage{amsmath}
\usepackage{graphicx}
\usepackage{amsfonts}
\usepackage{amssymb}%
\begin{document}

\title{A Critical Analysis of the Mean-Field Approximation for the Calculation of the
Magnetic Moment in the Friedel-Anderson Impurity\ Model}
\author{Gerd Bergmann\\Department of Physics\\University of Southern California\\Los Angeles, California 90089-0484\\e-mail: bergmann@usc.edu\\D:%
$\backslash$%
0aa%
$\backslash$%
Aa\_tex%
$\backslash$%
A\_paper%
$\backslash$%
Arb\_\_%
$\backslash$%
B152}
\date{\today}
\maketitle

\begin{abstract}
It is shown that the calculation of the magnetic moment of a Friedel-Anderson
impurity in mean-field theory is unreliable. A class of approximate solutions,
which contains the mean-field solution as an element, is expressed in rotated
Hilbert space and optimized. The optimal state has considerably lower energy
than the mean field solution and requires almost twice the Coulomb exchange
$U$ to become magnetic. Since most moment calculations of magnetic impurities,
for example the spin-density-functional theory, use the mean-field
approximation the resulting magnetic moments have to be critically reexamened.


\end{abstract}

The properties of magnetic impurities in a metal is one of the most
intensively studied problems in solid state physics. Although some of the
experimental anomalies were already discovered in the 1930's, it is still a
subject of great interest. The work of Friedel \cite{F28} and Anderson
\cite{A31} laid the foundation to understand why some transition metal
impurities form a magnetic local moment, while others don't. They considered a
host with an s-band in which a transition metal atom is dissolved. The
s-electrons can hop onto the d-impurity via the hopping matrix element
$V_{sd}$. The ten-fold degeneracy of a real d-impurity is simplified and
reduced to a two-fold degeneracy for spin up and spin down. If both states are
occupied they repel each other due to the Coulomb exchange energy. This yields
the Friedel-Anderson Hamiltonian
\begin{equation}
H_{FA}=\sum_{\sigma}\{\sum_{\nu=1}^{N}\varepsilon_{\nu}c_{\nu\sigma}^{\ast
}c_{\nu\sigma}+E_{d}d_{\sigma}^{\ast}d_{\sigma}+\sum_{\nu=1}^{N}V_{sd}%
(\nu)[d_{\sigma}^{\ast}c_{\nu\sigma}+c_{\nu\sigma}^{\ast}d_{\sigma}%
]\}+Un_{d+}n_{d-} \label{hfa0}%
\end{equation}
Here a finite s-band with $N$ states is used. The $c_{\nu\sigma}^{\ast}$ and
the $d_{\sigma}^{\ast}$ are the creation operators of the (free) s-electrons
and the d-impurity. The $d_{\sigma}^{\ast}$-states are assumed to be
orthogonal to the s-states $c_{\nu}^{\ast}$ (in the following I denote single
electron states by their creation operator).

In the limit of $V_{sd}=0$ and $E_{d}<\varepsilon_{F},E_{d}+U>\epsilon_{F}$
the d-impurity is magnetic. Anderson concluded that the magnetic moment
survives for small but finite $V_{sd}$ and derived the criteria for a magnetic
state and the size of the moment in a mean-field approximation. He found a
magnetic state if the product of $Ug_{d}>1$ where $g_{d}$ is the additional
density of states of the d-resonance.

Kondo \cite{K8} brought a new twist into the magnetic impurity problem when he
showed that multiple scattering of conduction electrons by a magnetic impurity
yields a divergent contribution to the resistance in perturbation theory. In
the following three decades a large number of sophisticated methods were
applied to better understand and solve the Kondo and Friedel-Anderson model,
and it was shown that at zero temperature a Friedel-Anderson impurity is in a
non-magnetic singlet state. However, above the Kondo temperature the impurity
shows a magnetic moment, and there is a great interest in the size of this moment.

There is a large body of \ research in which the magnetic moment of impurities
is calculated \cite{K46}, \cite{K47}, \cite{M41}, \cite{D28}, \cite{D33}.
Generally spin-density-functional theory is used for this task. Within this
theory the electronic structure of the host and the impurity is calculated
from first principles without any adjustable parameters. In particular the
strength of the Coulomb and exchange interaction are obtained from first
principles. However, in the final step the mean-field method is applied to
obtain the local magnetic moment. Although this is a zero-temperature
calculation (where the impurity should be in the Kondo singlet state) it is
generally argued that such a calculation yields the magnetic moment above the
Kondo temperature (which, at lower temperatures, is hidden in the singlet state).

In this letter I will show that the mean-field result for the magnetic moment
of impurities is not reliable. By rewriting the mean-field solution in a
rotated basis and optimizing the solution I obtain solutions which are much
lower in energy, require a much larger critical $U$ for the formation of a
moment and yield smaller moments. And this despite the fact that the improved
solution has the same structure (in the rotated basis) as the mean-field
solution. Since there is a large body of spin-density-functional theory
calculations for magnetic impurities, a reevaluation of this method might be required.

I start with Anderson's (potentially) magnetic state which he obtained as a
mean-field solution. Anderson replaced the Hamiltonian $H_{FA}$ by%

\begin{align}
H_{mf}  &  =H_{F+}+H_{F-}-U\langle n_{d+}\rangle\langle n_{d-}\rangle
\label{hmf1}\\
H_{F\sigma}  &  =\Sigma_{\nu}\varepsilon_{\nu}c_{\nu\sigma}^{\ast}c_{\nu
\sigma}+\sum_{\nu=1}^{N}V_{sd}(\nu)[d_{\sigma}^{\ast}c_{\nu\sigma}%
+c_{\nu\sigma}^{\ast}d_{\sigma}]+E_{d,\sigma}d_{\sigma}^{\ast}d_{\sigma}
\label{hfr}%
\end{align}
where $\langle n_{d+}\rangle$ and $\langle n_{d-}\rangle$ are the average
occupation numbers of the states $d_{+}^{\ast}$ and $d_{-}^{\ast}$ and
$E_{d,\sigma}=\left(  E_{d}+U\langle n_{d,-\sigma}\rangle\right)  $. The
solution of the mean-field method requires the diagonalization of two Friedel
resonance Hamiltonians $H_{F\sigma}$ with self consistent values for $\langle
n_{d+}\rangle$ and $\langle n_{d-}\rangle$. This straight-forward numerical
calculation yields the mean-field ground-state energy $E_{mf}$ and the
magnetic moment $\mu_{mf}$. The energy of the bare magnetic state $E_{b.m.}$
is subtracted from $E_{mf}$ where%

\begin{equation}
E_{b.m.}=2\sum_{\nu=1}^{n}\varepsilon_{\nu}+E_{d}-\varepsilon_{n} \label{eqv}%
\end{equation}
is the ground-state energy for $V_{sd}=0$ and $E_{d}<$ $\varepsilon_{F}%
,E_{d}+U$ $>\varepsilon_{F}.$

For the numerical calculation an s-band with a constant density of states is
used, ranging from $-1$ to $+1$. This band is divided into $N=48$ equal cell.
Each s-sub-band is half filled, i.e. the number of occupied states in each
spin sub-band is $n=N/2.$In Fig.1 the numerical results for $E_{mf}-E_{b.m.}$
are plotted for $\left\vert V_{sd}\right\vert ^{2}=0.05$. The Coulomb
repulsion $U$ is varied between 0.2 and 1.2. Together with the Coulomb
repulsion the $d^{\ast}$-state energy $E_{d}$ is varied so that $E_{d}$ and
$(E_{d}+U)$ lie symmetrically about the Fermi energy, i.e. $E_{d}=-{\frac
{1}{2}}U$.

In the mean-field calculation the impurity is non-magnetic for $U<U_{cr}%
\thickapprox0.275$. For $U>U_{cr}$ the spin up and down sub-bands split. The
resulting magnetic moments are plotted in Fig.2 (curve with circles).%

\[%
\raisebox{0.0083in}{\includegraphics[
height=3.731in,
width=4.2997in
]%
{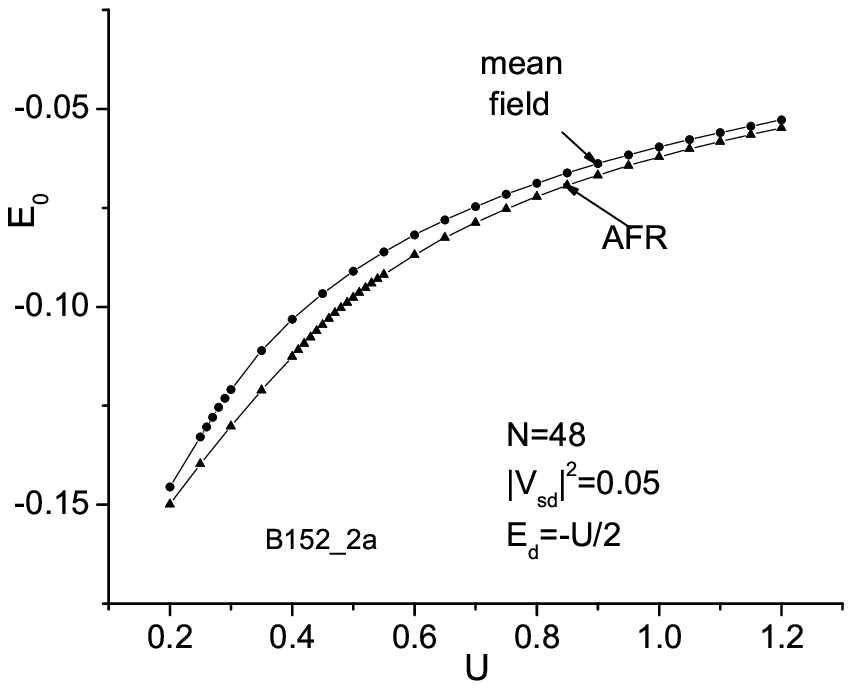}%
}%
\]
Fig.1: A comparison between the ground-state energies of the mean-field
calculation and the AFR method (see text below).%
\[
\]

Since the mean-field solution is the product of two $n$-electron states of the
two Friedel Hamiltonians $F_{F+}$ and $H_{F-}$ we consider these solutions of
the Friedel Hamiltonian (\ref{hfr}) in some detail. As shown in ref.
\cite{B91}, \cite{B92} the exact ground state of $H_{F}$ with $n$ (spinless)
electrons can be written in the form%

\begin{equation}
\Psi_{F}=\left[  A^{\prime}a_{0}^{\ast}+B^{\prime}d^{\ast}\right]  \prod
_{i=1}^{n-1}a_{i}^{\ast}\Phi_{0} \label{yfr}%
\end{equation}

Here $\Phi_{0}$ is the vacuum state and $a_{0}^{\ast}$ is a sister state to
$d^{\ast}$ which is built from the states of the s-band%

\begin{equation}
a_{0}^{\ast}=\sum_{\nu=1}^{N}\alpha_{\nu}^{0}c_{\nu}^{\ast} \label{afr}%
\end{equation}
Ref. \cite{B92} shows how to calculate the coefficients $\alpha_{\nu}^{0}$
from the parameters of the Hamiltonian $H_{F}$ and the occupation number $n$.
The $a_{i}^{\ast}$ are orthogonal to $a_{0}^{\ast}$ and to each other and
their $(N-1)$ sub-matrix of the s-band Hamiltonian $H_{0}=\sum\varepsilon
_{\nu}n_{\nu}$ is diagonal (see equ. \ref{h0'}). The states $a_{i}^{\ast}$ are
uniquely determined from the state $a_{0}^{\ast}$. Their form is
\begin{equation}
a_{i}^{\ast}=\sum_{\nu=1}^{N}\alpha_{\nu}^{i}c_{\nu}^{\ast}%
\end{equation}
The $a_{i}^{\ast}$ ($1\leq i\leq N-1$) together with $a_{0}^{\ast}$ represent
a new basis.

In this new basis the free electron Hamiltonian $H_{0}=\sum_{\nu=1}%
^{N}\varepsilon_{\nu}c_{\nu}^{\ast}c_{\nu}$ takes the form%
\begin{equation}
H_{0}=\sum_{i=1}^{N-1}E\left(  i\right)  a_{i}^{\ast}a_{i}+E\left(  0\right)
a_{0}^{\ast}a_{0}+\sum_{i=1}^{N-1}V_{fr}^{a}\left(  i\right)  \left[
a_{0}^{\ast}a_{i}+a_{i}^{\ast}a_{0}\right]  \label{h0'}%
\end{equation}
In the Hamiltonian ($\ref{h0'}$) the $a_{0}^{\ast}$-state represents an
artificial resonance state. I will call it in honor of Friedel an Artificial
Friedel Resonance state (AFR state). It is a sister state to the state
$d^{\ast}.$

The full (spin independent) Friedel Hamiltonian can be written as%

\begin{align}
H_{F}  &  =\sum_{i=1}^{N-1}E\left(  i\right)  a_{i}^{\ast}a_{i}+E\left(
0\right)  a_{0}^{\ast}a_{0}+E_{d}d^{\ast}d+V_{sd}^{a}(0)[d^{\ast}a_{0}%
+a_{0}^{\ast}d]\label{hfr'}\\
&  +\sum_{i=1}^{N-1}V_{sd}^{a}\left(  i\right)  \left[  d^{\ast}a_{i}%
+a_{i}^{\ast}d\right]  +\sum_{i=1}^{N-1}V_{fr}^{a}\left(  i\right)  \left[
a_{0}^{\ast}a_{i}+a_{i}^{\ast}a_{0}\right] \nonumber
\end{align}
where%

\begin{align}
E\left(  i\right)   &  =\sum_{\nu}\alpha_{\nu}^{i}\varepsilon_{\nu}\alpha
_{\nu}^{i}\nonumber\\
E\left(  0\right)   &  =\sum_{\nu}\alpha_{\nu}^{0}\varepsilon_{\nu}\alpha
_{\nu}^{0}\label{efr}\\
V_{sd}^{a}\left(  i\right)   &  =\sum_{\nu}V_{sd}\left(  \nu\right)
\alpha_{\nu}^{i}\nonumber\\
V_{fr}^{a}\left(  i\right)   &  =\sum_{\nu}\alpha_{\nu}^{i}\varepsilon_{\nu
}\alpha_{\nu}^{0}\nonumber
\end{align}

In the Hamiltonian ($\ref{hfr'}$) the $d^{\ast}$-state and the localized
$a_{0}^{\ast}$-state are on equal footing. The second line in equ.
($\ref{hfr'}$) yields the hopping between $a_{i}^{\ast}$ and $d^{\ast}$ (first
term) and $a_{i}^{\ast}$ and $a_{0}^{\ast}$ (second term). For the state
$\left(  A^{\prime}a_{0}^{\ast}+B^{\prime}d^{\ast}\right)  $ the individual
hopping matrix elements cancel each other, making $\Psi_{Fr}$ the ground state.

In the next step the mean-field solution is rewritten in the AFR-form of the
Friedel ground state. Since the Hamiltonian consists of a Friedel Hamiltonian
for each spin the mean-field state is the product of two states of the form of
equ.(\ref{yfr}). Therefore this mean-field state (the exact solution of the
mean-field Hamiltonian) can be written as%

\begin{align}
\Psi_{0}  &  =\left[  A_{-}a_{0-}^{\ast}+B_{-}d_{-}^{\ast}\right]  \left[
A_{+}a_{0+}^{\ast}+B_{+}d_{+}^{\ast}\right]  \prod_{\sigma,i=1}^{n-1}%
a_{i\sigma}^{\ast}\Phi_{0}\nonumber\\
\  &  =\left[  Aa_{0-}^{\ast}a_{0+}^{\ast}+Bd_{-}^{\ast}a_{0+}^{\ast}%
+Ca_{0-}^{\ast}d_{+}^{\ast}+Dd_{-}^{\ast}d_{+}^{\ast}\right]  \prod
_{\sigma,i=1}^{n-1}a_{i\sigma}^{\ast}\Phi_{0}\label{y0}\\
&  =A\Psi_{A}+B\Psi_{B}+C\Psi_{C}+D\Psi_{D}\nonumber
\end{align}
where%

\begin{equation}%
\begin{array}
[c]{ccccccc}%
A_{+}^{2}+B_{+}^{2} & = & 1 & , & A_{-}^{2}+B_{-}^{2} & = & 1\\
A & = & A_{+}A_{-} & , & B & = & A_{+}B_{-}\\
C & = & A_{-}B_{+} & , & D & = & B_{+}B_{-\,}%
\end{array}
\label{no0}%
\end{equation}
Each of the four states $\Psi_{A}$, $\Psi_{B}$, $\Psi_{C}$ and $\Psi_{D}$ is
normalized, and they are all orthogonal to each other. In the magnetic
solution one has $A_{+}\neq A_{-}$ and $B_{+}\neq B_{-}$. Also the two rotated
bases $\left\{  a_{0+}^{\ast},a_{i+}^{\ast}\right\}  $ and $\left\{
a_{0-}^{\ast},a_{i-}^{\ast}\right\}  $ are different in the magnetic state.

So far the many electron state in equ. (\ref{yfr}) is the mean-field solution.
This state consists of an electron background $\prod_{\sigma,i=1}%
^{n-1}a_{i\sigma}^{\ast}\Phi_{0}$ multiplied with the sum of four two-electron
states, consisting of the combinations $\left[  a_{0-}^{\ast}a_{0+}^{\ast
},d_{-}^{\ast}a_{0+}^{\ast},a_{0-}^{\ast}d_{+}^{\ast},d_{-}^{\ast}d_{+}^{\ast
}\right]  $ which have $S_{z}=0$. The mean-field wave function opens a new
playing field for variation to find the optimal state. One can optimize the
coefficients $A$, $B$, $C$ and $D$ while dropping the individual normalization
conditions (\ref{no0}) and replacing them by%

\begin{equation}
A^{2}+B^{2}+C^{2}+D^{2}=1 \label{no1}%
\end{equation}
Far more important one can optimize the states $a_{0+}$ and $a_{0-}$.

For this purpose the Hamiltonian $H_{FA}$ is expressed in the bases $\left\{
a_{0+}^{\ast},a_{i+}^{\ast}\right\}  $ and $\left\{  a_{0-}^{\ast}%
,a_{i-}^{\ast}\right\}  .$ One obtains for the expectation value of the
ground-state energy $E_{0}$%

\begin{align}
E_{0}  &  =A^{2}\left[  E_{-}\left(  0\right)  +E_{+}\left(  0\right)
\right]  +B^{2}\left[  E_{-}\left(  0\right)  +E_{d}\right]  +C^{2}\left[
E_{+}\left(  0\right)  +E_{d}\right]  +D^{2}\left[  2E_{d}+U\right]
\nonumber\\
&  +2\left(  AB+CD\right)  V_{sd}^{-}\left(  0\right)  +2\left(  AC+BD\right)
V_{sd}^{+}\left(  0\right)  +\sum_{\sigma,i=1}^{n-1}E_{\sigma}\left(
i\right)  \label{e0}%
\end{align}

For a given set of states $\left\{  a_{0\pm}^{\ast},a_{i\pm}^{\ast}\right\}  $
the energy $E_{0}$ in eq. (\ref{e0}) depends on the coefficients $A$, $B$, $C$
and $D$. One obtains the lowest energy by varying $E_{0}$ with respect to
these coefficients. This yields a 4x4 matrix for the coefficients vector
$(A,B,C,D)$. The lowest eigenvalue gives the energy expectation value, and its
eigenvector gives the coefficients. The resulting state I denote as the
magnetic state $\Psi_{AFR}$ and the solution as the AFR solution.

The central part of the numerical calculation is the variation of the states
$a_{0+}^{\ast}$ and $a_{0-}^{\ast}$ until the absolute minium of the energy is reached.

As in the mean-field theory the numerical calculation itself determines
whether the lowest state possesses a magnetic moment or not. If the solution
is magnetic then $a_{0+}^{\ast}$ and $a_{0-}^{\ast}$ approach different states
and the coefficients B and C have different values. The resulting magnetic
moment is defined as the difference in the occupation of the $d_{+}^{\ast}$
and $d_{-}^{\ast}$ states, i.e. $\mu=B^{2}-C^{2}$.

In Fig.1 the energy expectation value $E_{0}$ of the optimal magnetic state
$\Psi_{AFR}$ is plotted as the curve with the triangles (again the same energy
$E_{b.m.}$ has been subtracted). The new ground-state energy lies considerably
below the mean-field energy.

In Fig.2 the resulting magnetic moments that one obtains with the mean-field
approximation and with the new method are plotted. One recognizes that the new
solution suppresses the magnetic moment up to a considerably larger value of
$U_{cr}\approx0.46$. This is almost twice the value of the mean-field theory.%
\[%
{\includegraphics[
height=3.2744in,
width=3.9219in
]%
{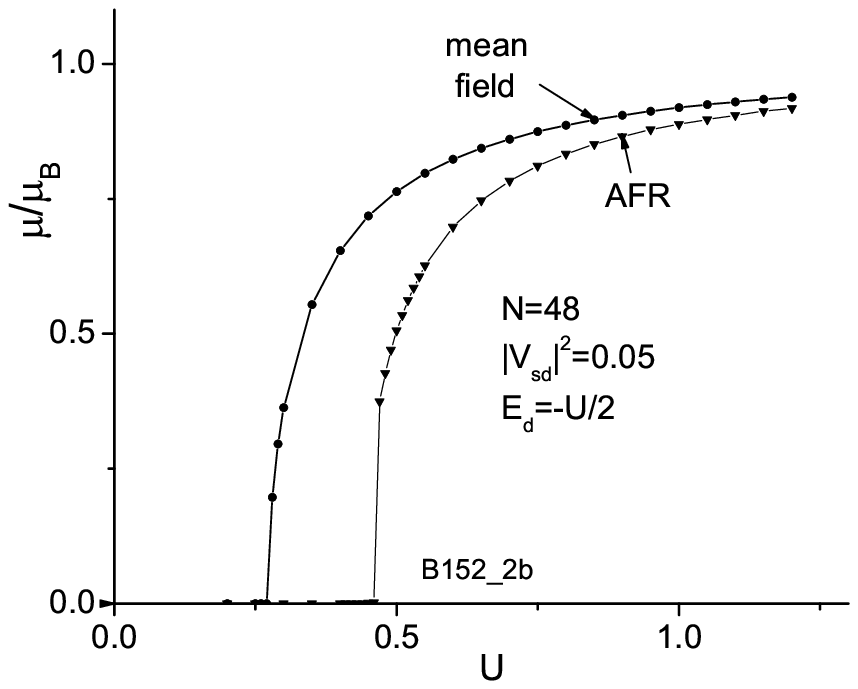}%
}%
\]
Fig.2: The magnetic moment as a function of the Coulomb energy $U,$ using the
mean-field solution and the AFR-method of the present calculation.\newline%
\[
\]

How do we have to interpret the fact that the AFR solution suppresses the
magnetic moment up to a much larger critical Coulomb exchange interaction?
Since this state has a lower energy expectation value, does this mean that its
magnetic moment is more reliable? The author prefers a somewhat different
interpretation. The Friedel-Anderson impurity does not like a broken symmetry.
The mean-field approach does not give the multi-electron state any wiggling
room. Only the values for $E_{d,\sigma}$ can be altered with increasing $U$.
The AFR solution on the other hand possesses a lot more flexibility since the
AFR states can adjust. Therefore the symmetric multi-electron state survives
to a considerably larger Coulomb exchange interaction. It might be that
neither state yields the right magnetic moment for the impurity. The present
calculation raises serious questions about the mean-field approach. This might
also apply to the spin-density-functional theory (SDFT) for magnetic
impurities. This theory is a very complex theory and it is difficult to judge
from the outside all the intricacies. It should yield the correct charge and
spin densities for the correct functional. But in the final step the majority
of SDFT calculations use a two-spin-fluid model where each electron fluid
adjusts in the (mean) field of the other.

To conclude, in this paper an approximate solution for the Friedel-Anderson
impurity is expressed in a rotated Hilbert space $\left\{  a_{0\pm}^{\ast
},a_{i\pm}^{\ast}\right\}  $. Its center piece are two artificial resonance
states $a_{0+}^{\ast},a_{0-}^{\ast}$ for the spin up and down s-electrons.
They determine uniquely the remaining bases $\left\{  a_{i+}^{\ast}\right\}
$,$\left\{  a_{i-}^{\ast}\right\}  $. The AFR states are combined with the
d-electrons for spin up and down $d_{+}^{\ast},d_{-}^{\ast}$ into two-electron
states of total $S_{z}=0$, i.e. $\left[  Aa_{0-}^{\ast}a_{0+}^{\ast}%
+Bd_{-}^{\ast}a_{0+}^{\ast}+Ca_{0-}^{\ast}d_{+}^{\ast}+Dd_{-}^{\ast}%
d_{+}^{\ast}\right]  $. Then the $\left(  n-1\right)  $ lowest states of the
two $\left(  N-1\right)  $ bases $\left\{  a_{i\pm}^{\ast}\right\}  $ are
occupied yielding the s-electron background $%
{\textstyle\prod\limits_{i=1,\sigma}^{n-1}}
a_{i\sigma}^{\ast}\Phi_{0}$. The compositions of the AFR states $a_{0+}^{\ast
},a_{0-}^{\ast}$ are calculated by numerical variation which rotates the
s-electron bases in Hilbert space.

The energy of the resulting state lies clearly below the mean-field solution.
The critical value of the Coulomb exchange energy $U_{cr}$ for the formation
of a magnetic moment is almost twice as large as in the mean-field solution.
Since in many calculations of the magnetic moment of impurities the mean-field
approximation is used one has to reevaluate the resulting moments. This may
also apply to the impurity calculations which use the spin-density-functional
theory because in the majority of these calculations the mean-field theory is
used in the final analysis.

Since the ground state of the Friedel-Anderson impurity is a singlet state one
might suspect that the structure of the new solution with the lower energy and
smaller magnetic moment is somewhat closer to the singlet state than the
mean-field solution. This is not the case. Both the mean-field and the present
solution are in a symmetric state for small $U$; both show a similar asymmetry
between spin up and down in the magnetic state. The mean-field solution
belongs to the same class of wave functions as the here presented one (which
are given by the general form of equ. (\ref{y0})).


\begin{thebibliography}{99}                                                                                               %


\bibitem {F28}J.Friedel, Philos.Mag. 43, 153 (1952); Adv.Phys. 3, 446 (1954);
Philos.Mag.Suppl. 7, 446 (1954); Can.J.Phys. 34, 1190 (1956); Nuovo Cimento
Suppl. 7, 287 (1958); J. Phys.Radium 19, 38 (1958)\newline

\bibitem {A31}P.W.Anderson, Phys.Rev. 124, 41 (1961)\newline

\bibitem {K8}J.Kondo, Prog.Theor.Phys. 32, 37 (1964) \newline

\bibitem {K46}S.K.Kwon and B.I.Min, Phys.Rev.Lett. 84, 3970 (2000)\newline

\bibitem {K47}R.B.Sahu and L.Kleinman, Phys.Rev. B67, 094424 (2003)\newline

\bibitem {M41}M.E.McHenry, J.M.MacLaren, D.D.VVendensky, M.E.Eberhart and
M.L.Prueitt, Phys.Rev. B40, 10111 (1989)\newline

\bibitem {D28}R.Podloucky, R.Zeller and P.H.Dederichs, Phys.Rev. B22, 5777
(1980)\newline

\bibitem {D33}V.I.Anisimov and P.H.Dederichs, Solid State Commun., 84, 241
(1992)\newline

\bibitem {B91}G.Bergmann, Z.Physik B102, 381 (1997)\newline

\bibitem {B92}G.Bergmann, Eur.Phys.J.B2, 233 (1998)\newline
\end{thebibliography}
\end{document}